\documentclass[]{IEEEtran}

\usepackage{graphicx}
\usepackage{url}

\begin{document}

\title{\LARGE Measurement of Radio-Frequency Radiation Pressure}



%
\author{\IEEEauthorblockN{Alexandra Artusio-Glimpse,
Matt T. Simons,
Ivan Ryger,
Marc Kautz,
John Lehman,
 and
Christopher~L.~Holloway} \\
National Institute of Standards and Technology (NIST), Boulder,~CO~80305,  holloway@boulder.nist.gov\\}


\maketitle

\begin{abstract}
We perform measurements of the radiation pressure of a radio-frequency (RF) electromagnetic field which may lead to a new SI-traceable power calibration.
There are several groups around the world investigating methods to perform more direct SI traceable measurement of RF power (where RF is defined to range from 100s of MHz to THz).  A measurement of radiation pressure offers the possibility for a power measure traceable to the kilogram and to Planck's constant through the redefined SI. Towards this goal, we demonstrate the ability to measure the radiation pressure/force carried in a field at 15~GHz.
\end{abstract}

%
\IEEEpeerreviewmaketitle

\pagenumbering{gobble}

\vspace{-1mm}

\section{Introduction}
The current method of power traceability is typically based on an indirect path through a thermal measurement using a calorimeter. The world of measurement science is changing rapidly with the SI redefinition planned for 2018. As a result of the shift towards fundamental physical constants, the role of primary standards must change. This includes radio-frequency (RF) power, which is currently traceable to electrical units through thermal detectors. Various groups are investigating different approaches to perform more direct SI traceable power measurements.  This includes measurements based on Rydberg atoms \cite{r1}-\cite{r3} and measurement of power through the radiation pressure carried by an electromagnetic (EM) field \cite{p1}, \cite{p2}. The Rydberg atom approach allows a direct SI traceable measurement through Planck's constant, while the radiation pressure approach allows for a SI traceable measurement through the kilogram. However, with the redefinition of the SI in 2018, the kilogram will be traceable directly to Planck's constant \cite{si}.

While measurement of radiation pressure has been demonstrated for high laser power, such radiation pressure measurements have not been demonstrated at RF [here RF refers to frequencies ranging 100’s MHz to 100’s of GHz, and to just below THz].  The concept of measuring radiation pressure stems from the fact that EM fields carry a momentum as they propagate through space and the momentum results in an EM pressure expressed as (in units of N/m$^2$) \cite{marion}
\begin{equation}
{\bf {\cal P}}=\frac{\langle{\bf E}\times{\bf H}\rangle}{c} \,\,\, ,
\label{pp}
\end{equation}
where $c$ is the speed of light {\it in vacuo}, and  ${\bf E}$ and ${\bf H}$ are the electric and magnetic fields. By measuring this pressure (via a force measurement), we can obtain a measurement of the RF power carried in the RF field. Here we will discuss the set of experiments and the devices used to perform, to our knowledge, the first set of RF radiation pressure measurements.

\section{Experimental Setup and Pressure Sensor}

\begin{figure}
\center
\includegraphics[width=2.7in]{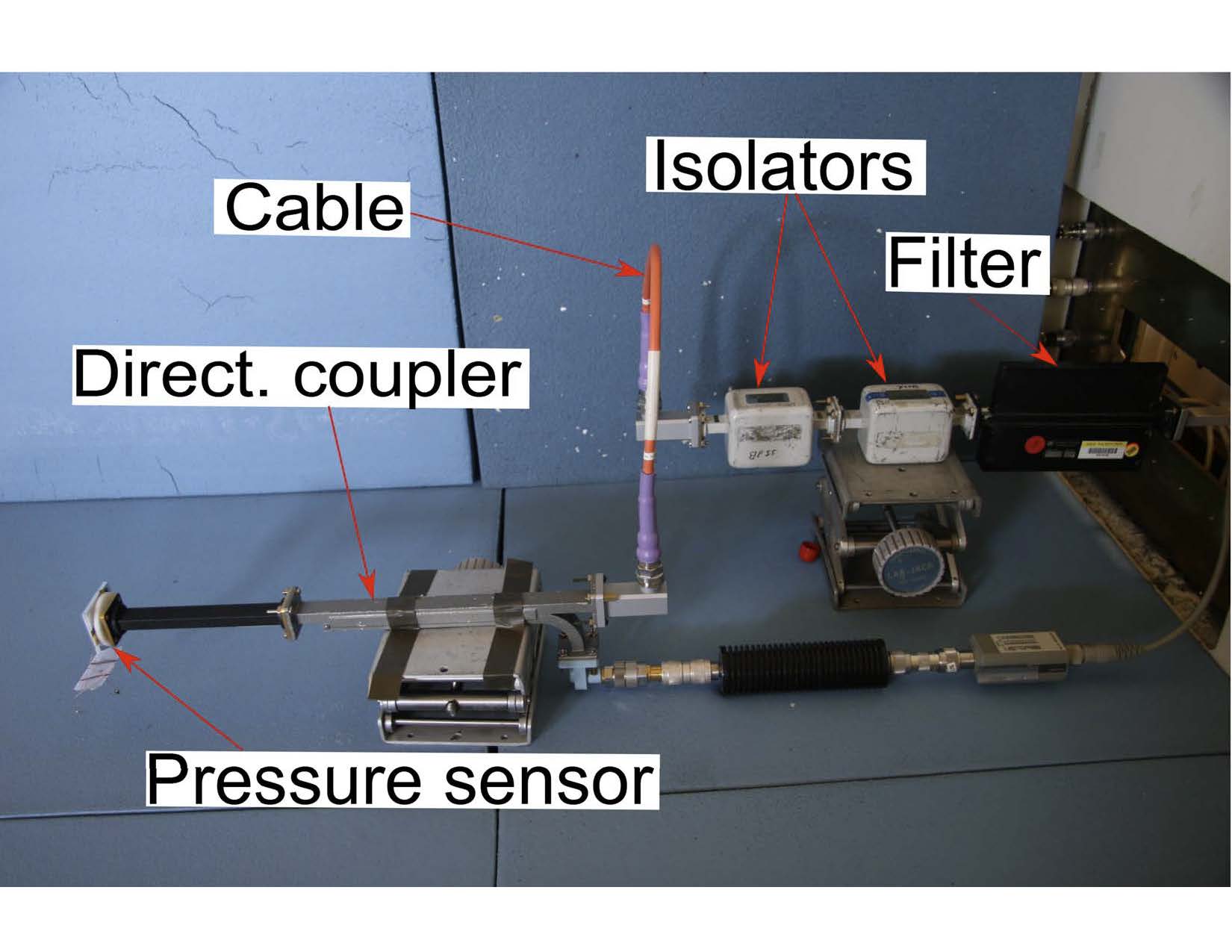}
\caption{Photo of the setup used for the RF radiation pressure experiments.}
\label{f1}
\end{figure}

A photo of the experimental setup is shown in Fig.~\ref{f1}.  A signal generator (SG) feeds a 200~W amplifier with a WR62 waveguide output. The output was run through a filter and two isolators and then to a waveguide/coax adapter. A 0.5~m cable was used to connect to a second waveguide/coax adapter to feed a second waveguide section (the cable was used to isolate the pressure sensor from mechanical vibrations caused by the amplifier). The waveguide was connected to a directional coupler and then to the last section of waveguide (the black one in the figure). The pressure sensor was connected to the output of this black section of waveguide.

\begin{figure}
\center
\includegraphics[width=2.6in]{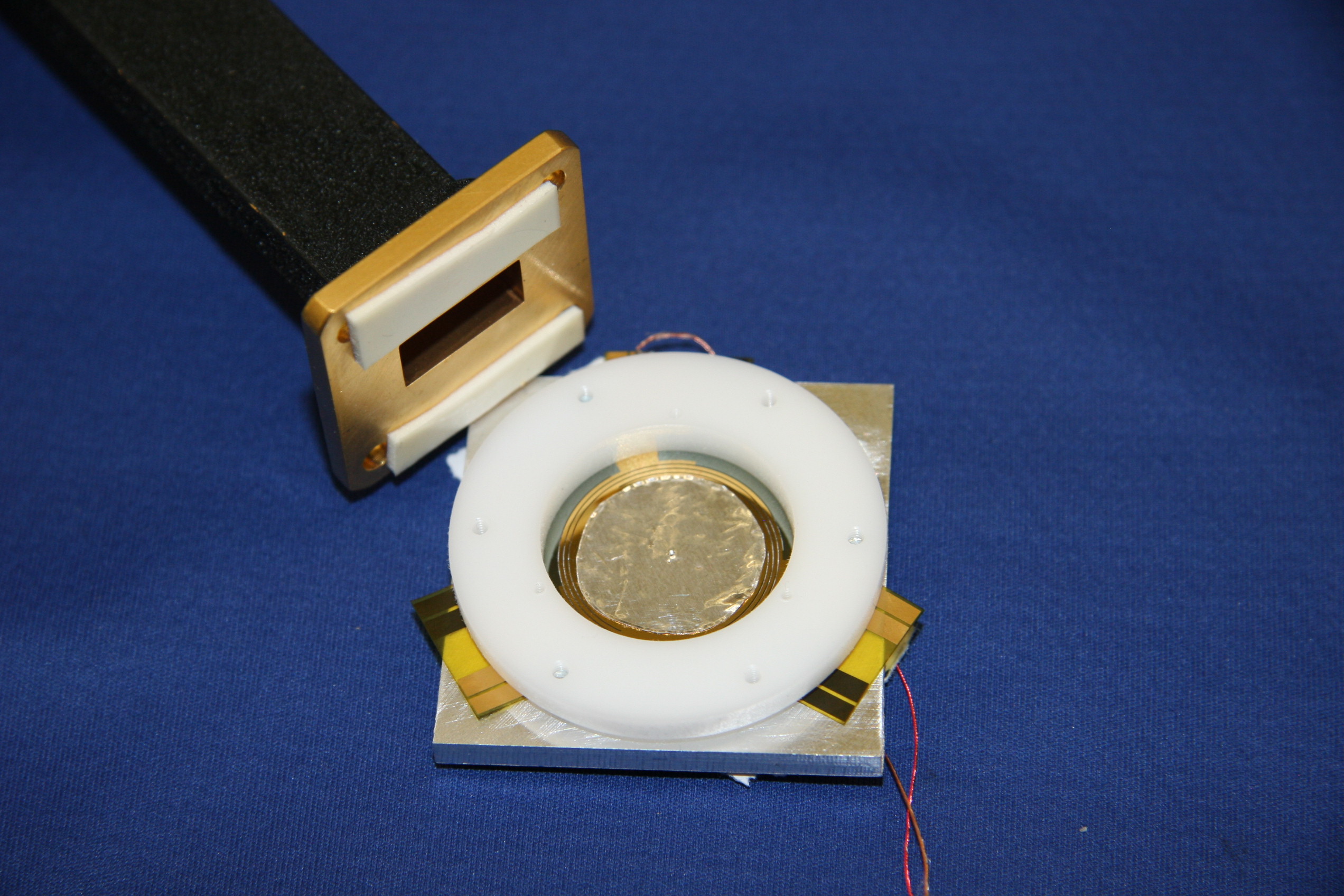}
\caption{Photo of the pressure sensor detached from the waveguide.}
\label{sensor}
\end{figure}

The radiation pressure device (shown in Fig. \ref{sensor}) is a capacitor-based force sensor. Upon reflection of the plane-wave RF beam normally incident on an aluminum reflector, a force given by the change in momentum of the reflected beam deflects a silicon spring. This changes the plate spacing of a parallel-plate capacitor, which sets a Wien capacitor bridge out of balance producing a voltage signal. In these experiments, we drove the bridge with a 20~kHz, 1~V sinusoidal source and demodulated the bridge signal with a lock-in amplifier locked to 20~kHz using a 30~ms, 6~dB lowpass filter for noise suppression. We recorded the output of the lock-in amplifier with an oscilloscope set to AC and triggered to the modulation of the RF source.

The spring itself is cut from crystalline silicon wafer. It includes a flat central disk, 20~mm in diameter, which supports the aluminum reflector on the front side (see Fig.~\ref{sensor}) and a gold electrode on the back side. The disk is surrounded with three Archimedean spiral legs that are deep etched through the silicon wafer, connecting the central disk to an annular support for mounting. Given normal deflection of the spring, where the central disk moves along its surface normal, the spring stiffness is approximately 170~N/m. The radiation force on the spring from the reflected RF beam is given by (in units of N):
\begin{equation}
F =  \frac{2P}{c}\left(R+(1-R)\frac{\alpha}{2}\right)
\label{e2}
\end{equation}
where $P$ (in units of W) is the RF power incident on the reflector, $R$ is the reflectance of the aluminum, and $\alpha$ is the fraction of non-reflected RF that is absorbed. In this experiment, we assume all of the non-reflected RF energy is absorbed by the spring, or $\alpha=1$.

The spring was clamped to a rigid aluminum back plate with a 30~$\mu$m thick polyimide spacer. With no force acting on the spring, the capacitance between the spring and aluminum back plate was 255~pF. Ferrite beads and a low pass filter were used to isolate the bridge electronics from any potential RF leakage, which added 24~pF of parasitic capacitance to the sensor. In this configuration, the signal sensitivity of the sensor was $-4.5$~V/pF. With a thermocouple attached to the aluminum back plate, we were able to track the sensor temperature throughout RF beam injection.

\section{Experimental Results}

\begin{figure}
\center
\includegraphics[width=3.5in]{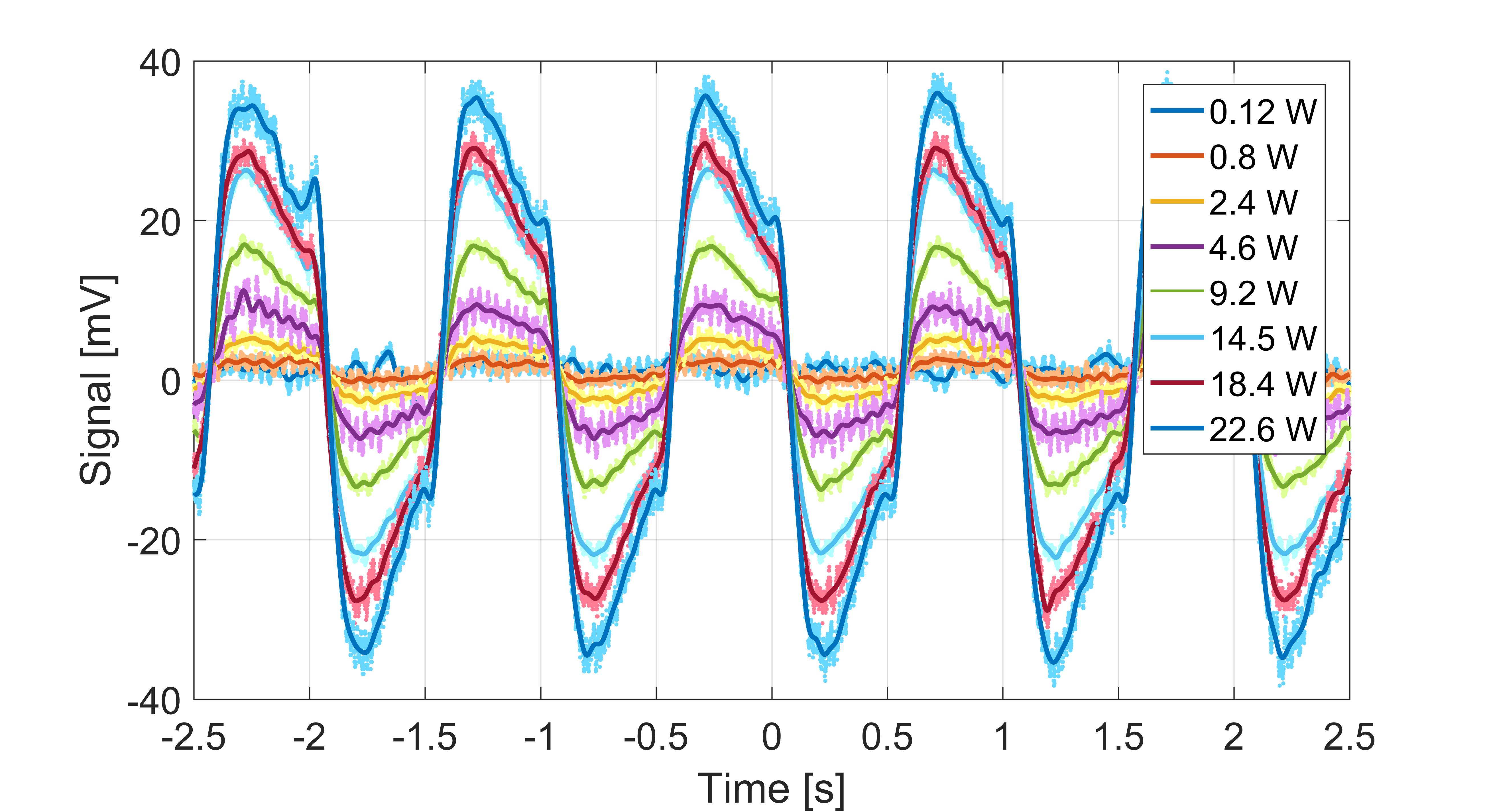}\\
\centering
\tiny{(a) Signal voltage versus time.}\\
\includegraphics[width=3.5in]{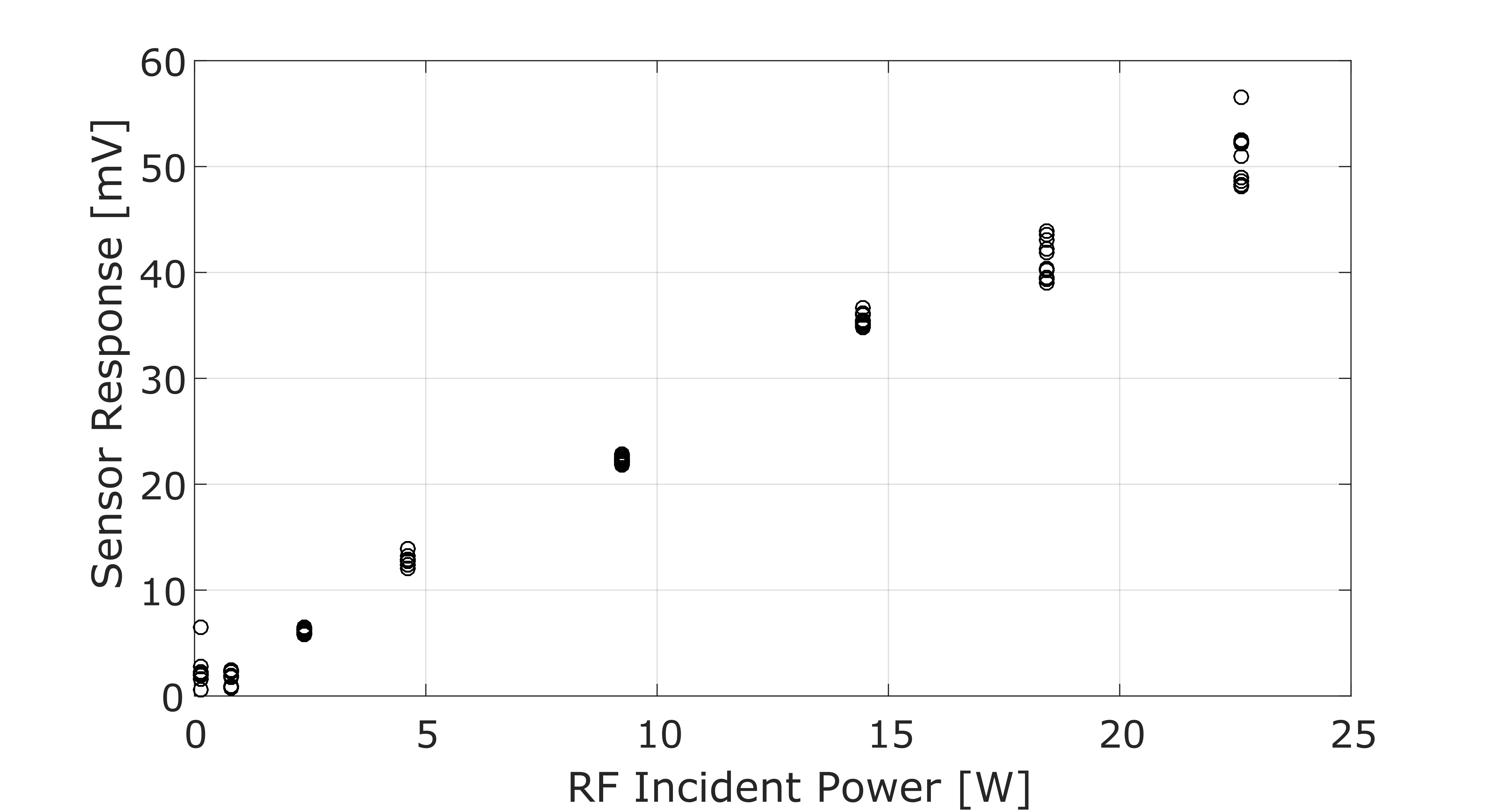}\\
\tiny{(b)  Signal voltage versus RF power.}\\
\caption{Signal voltage from the pressure sensor. The results in part (b) include repeated experiments.}
\label{f3}
\end{figure}

During the experiments, the output of the SG was varied such that the power (measured with a traditional power meter) at the output of the waveguide ranged from 0.12~W to 22.6~W. The RF pressure-sensor measured voltages for these different power levels are shown in Fig.~\ref{f3}. Fig.~\ref{f3}(a) shows the signal versus time, where the RF power was modulated at 1~Hz and Fig. \ref{f3}(b) shows the signal voltage of the sensor for different RF power levels.  The results in these figures illustrate that the radiated power can be detected with this pressure sensor.

In order to estimate the value of forces that the pressure sensor was detecting, we used eq.~(\ref{e2}) and assumed a reflectance of $R=0.90$. For the power range of 1~W to 23~W, we calculated the force range to be 6.3~nN to 145.6~nN. For reference, a grain of maize pollen weighs approximately 2.5~nN; a RDA of vitamin B12 weighs approximately 23.5~nN; a human eyelash weighs approximately 686~nN; and a fruit fly weighs approximately 1960~nN. Thus, we can detect a force that is equivalent to a grain of maize pollen or 300 times smaller than that of a fruit fly.

\vspace{-2mm}

\section{Conclusions}

We have demonstrated for the first time that it is possible to detect the radiation pressure carried by an RF field. We illustrated this by performing experiments at 15~GHz, where a pressure sensor was placed at the end of a waveguide.   These types of power measurements could lead to a fundamentally new approach for calibrating RF power and lead to a new SI traceable approach for RF power.

In future work, we will perform similar experiments at different RF frequencies, and we will calibrate the pressure sensor in order to give a direct measurement of the RF power carried in the EM field. Furthermore, this approach has the capability of measuring power levels well above kWs. However, we hope to develop pressure sensors to measure much smaller power levels ($<$1~mW, if possible).  Finally, the uncertainties of these types of measurements are currently being investigated, and it is believed that they can be made much smaller than current techniques.


\vspace{-2mm}

\begin{thebibliography}{1}

\bibitem{r1} C.L. Holloway, M.T. Simons, J.A. Gordon, P.F. Wilson, C.M. Cooke, D.A. Anderson, and G. Raithel,
     {\it IEEE Trans. on Electromagnetic Compat.,}
{\bf 59}(2), 717-728, 2017.

\bibitem{r2} J.A. Sedlacek, A. Schwettmann, H. K\"{u}bler, R. L\"{o}w, T. Pfau and J.P. Shaffer,
    {\it Nature Phys.}, {\bf 8}, 819, 2012.


\bibitem{r3} C.L. Holloway, J.A. Gordon, A. Schwarzkopf, D.A. Anderson, S.A. Miller, N. Thaicharoen, and G. Raithel, 
{\it IEEE Trans. on Antenna and Propag.}, {\bf 62}(12), 6169-6182, 2014.

\bibitem{p1} I. Ryger, A. Artusio-Glimpse, P. Williams, N. Tomlin,
MI. Stephens, K. Rogers, M. Spidell, J Lehman,
``Micromachined force balance for optical power measurement by radiation pressure sensing''
submitted, 2018.

\bibitem{p2} P.A. Williams, J.A. Hadler, R. Lee, F.C. Maring, and J.H. Lehman,
    {\it Opt. Lett.}, vol. 38, no. 20, 4248–4251, 2013.

\bibitem{si} ``Redefining the kilogram'', \newline https://www.nist.gov/pml/productsservices/redefining-kilogram,  2018.

\bibitem{marion} J.B. Marion and M.A. Heald,
{\bf Classical Electromangetic Radiation}, {\it 2nd Edition}. Academic Press, 1980.

\end{thebibliography}
\end{document}